\documentstyle[aps,prl,graphicx]{revtex}

\begin{document}
\title{Study of the $\pi$N and $\eta$p de-excitation channels of the
N$^{*}$ and $\Delta$ baryonic resonances between 1470~MeV and 1680~MeV.}

\author{
B. Tatischeff$^{1}$\thanks{E-mail : tati@ipno.in2p3.fr},
J. Yonnet$^{1}$, M. Boivin$^{2}$, M. P. Comets$^{1}$, P. Courtat$^{1}$,\\
R. Gacougnolle$^{1}$, Y. Le Bornec$^{1}$, E. Loireleux$^{1}$, M.
MacCormick$^{1}$, and F. Reide$^{1}$}

\address{
$^1$Institut de Physique Nucl\'eaire, Universit\'e Paris-Sud, CNRS/IN2P3,
F--91406 Orsay Cedex, France\\
$^2$Laboratoire National Saturne, CNRS/IN2P3, F--91191 Gif-sur-Yvette
Cedex, France}

\maketitle
\vspace*{1cm}
\begin{abstract}
Two reactions: pp$\to$ppX and pp$\to$p$\pi^{+}$X are used in order to study
the 1.47$\le$M$\le$1.680~GeV baryonic mass range. Three different final
states are considered in the invariant masses: N$^{*}$ or $\Delta^{+}$,
n$\pi^{+}$ or p$\pi^{0}$, and p$\eta$. The two last channels are defined by
software cuts applied to the missing mass of the first reaction.
Several narrow structures are extracted with widths $\sigma(\Gamma)$
varying between 3 and 9~MeV. Some structures are observed in one channel but
not in another. Such non-observation may be due either to the spectrometer 
momenta limits, or to the physics (for example no such disintegration
channel allowed from the narrow state discussed).\\
\hspace*{4.mm}We tentatively conclude
that the broad PDG baryonic resonances: N(1520)D$_{13}$, N(1535)S$_{11}$,
$\Delta$(1600)P$_{33}$, N(1650)S$_{11}$, and N(1675)D$_{15}$ are collective
states built from several narrow and weakly excited resonances, each having
a (much) smaller width than the one reported by PDG.
\end{abstract}

PACS numbers: 13.30.Eg, 14.20.Gk

\section{Introduction}
\hspace*{4.mm}For a number of years partial-wave analysis of baryonic
resonances have been extensively studied using mainly
$\pi$N$\to\pi$N, $\pi$N$\to\pi\pi$N,
and $\gamma$N$\to\pi$N reactions \cite{pdg}. The study of the dynamics of the
$\pi$N$\to\pi\pi$N reaction \cite{kam} is also useful for improving our
knowledge about
meson production, as well as our knowledge of the baryonic resonances.
 These resonances have also been 
studied with pp$\to$pN$^{*}$ reactions \cite {ede}. The dispersion of
the masses found by different experiments, remains generally less
than 20~MeV. However the widths of these resonances,
as reported by the various authors, differ by a large amount. For
example, the width of the N(1440)P$_{11}$ resonance, extracted from the
 $\pi$N$\to\pi$N reaction, varies between a lower limit of
135$\pm$10~MeV and an upper limit of 545$\pm$170~MeV \cite{pdg}. Moreover,
the many theoretical
width attributions inside constituent quark models \cite{cap} are often
much smaller than those reported by the experiments.
Such discrepancies between different experimental results and between 
calculated and experimental widths have sometimes been emphasized, and
have been used for example to test \cite{iac} whether
 the N(1535)S$_{11}$ could be a pentaquark $q^{4}\bar{q}$ state.
It appears therefore that a study of ``low mass'' baryon resonances with
more precise experiments than those previously performed - with a
good resolution,
small binning, and large statistics - will be a sensitive test for the many
existing models. Another important test to select among the models, would be
to measure, with a high precision, the
branching ratios for various de-excitation modes.  In this respect, a recent
theoretical work \cite{kis} finds
a B$_{N\pi}$ ratio for the S$_{11}$(1535) varying from 29$\pm$1$\%$
up to 67$\pm$1$\%$ depending on the model considered. This question was
also studied in different theoretical papers, with the emphasis on positive
and negative parity resonances \cite{bro}, or J$^{\pi}$=3/2$^{-}$ resonances
\cite{kol}.\\ 
\hspace*{4.mm}If we exclude some recent experiments performed at JLab, 
nearly all previous studies in the non-strange sector, were done without any
attempt to have a good resolution, and therefore a suitable binning. Now,
it is possible to do measurements with a better resolution, and finer binning,
than before. Such an experiment was performed at the now defunct Saturne
synchrotron, on the SPES3 beam line.\\
\hspace*{4.mm}In this paper the data from the
p~p~$\rightarrow$~p~p~X and p~p$\rightarrow$~p~$\pi^{+}$~X reactions
studied at two incident proton energies: T$_{p}$=1520~MeV and
T$_{p}$=1805~MeV are reported. The incident energies open up
 the $\pi^{0}$p ($\pi^{+}$n) and $\eta$p de-excitation
channels of
the N$^{*}$ and $\Delta^{+}$ baryonic resonances. For the first reaction,
p~p~$\rightarrow$~p~p~X, mass
ranges 1470$\le M_{pX}\le$1585~MeV using the lowest incident
energy and 1560$\le M_{pX}\le$1680~MeV using the highest incident energy were
studied. The second reaction, p~p$\rightarrow$~p~$\pi^{+}$~X, allows the
study of the n$\pi^{+}$ de-excitation
channel of the N$^{*}$ and $\Delta^{+}$ baryonic resonances in the range
1470$\le M_{n\pi^{+}}\le$1640~MeV and with a slight spectrometer
angle dependence.\\
\hspace*{4.mm}The study of nucleon resonances is clearly connected with the
production of mesons either close to threshold or at higher energies. A great
number of papers are related to such studies and all these papers cannot
be quoted here \cite{col}, \cite{kamf}.\\
\hspace*{4.mm}In the baryonic mass range studied in this paper, one major
problem is to distinguish between genuine quark model states and exotic
states. Unlike the situation at low baryonic masses, where new narrow states
are clearly exotic, for the mass range discussed in this work the answer
is not so simple, since, as
already stated, the calculated widths of quark model states are sometimes
much smaller than the experimental widths reported by PDG. The situation is
further complicated by the possibility of connecting new baryons to the
``missing resonances'' predicted by the constituent quark model, but,
so far, not observed.
Clearly, measurements of as many properties as possible
of the baryonic resonance excitations and disintegrations, would be useful.\\ 
\hspace*{4.mm}This paper is constructed in the following way: the main
properties of the experimental layout are briefly recalled in the second
section. The third section describes the analysis, and details the various
intermediate results for one given angle and one given missing mass
selection. The different ways of subtracting
events giving the experimental signal but corresponding to the
``uncorrelated'' missing mass
contribution (hereafter called the physical background), will
be considered. Results are presented and discussed in the fourth section, 
and are compared with many theoretical quark model results.
A general discussion then follows and the data for several de-excitation
branches are compared. The last section contains a summary of the results and a 
conclusion.\\
\hspace*{4.mm}Our results on narrow baryonic resonances in the mass range
1.0$\le$M$\le$1.46~GeV were already published in \cite{bor1}, and those in the
mass range 1.72$\le$M$\le$1.79~GeV were previously published in \cite{bor3}. 
\section{Properties of the magnet and detector layout.}
A complete description was already published in a previous paper \cite{bor1}. 
This desription included a presentation
of the experimental set up, as well as the performances of the experimental
apparatus, the checks performed,
the simulation code used to evaluate raw data normalization,
and the various normalizations which had to be introduced. Such a lengthy
description  will not be repeated here, only the main properties
will be given.\\
\hspace*{4.mm}The measurements were performed  at the Saturne synchrotron
beam facility. The experimental signature required the simultaneous
detection of two positively charged particles, using the SPES3 spectrometer
and detection system (see Fig.~1). A liquid H$_{2}$ target was used.
The spectrometer allowed a large momentum range of the detected particles:
600~MeV$<$pc$<$1400~MeV at B=3.07~T, with help of several
drift chambers. The trigger consisted of four planes of plastic
scintillator hodoscopes. Particles were identified mainly by their time of
flight. The particle momenta were determined using one MIT
and several CERN type drift chambers. The resolution in the invariant mass
spectrum was measured using the final state interaction peak (FSI) from
the p~p~$\to$~p~p~X reaction. It was close to $\sigma(\Gamma) \approx$~0.9~MeV.
The resolution of the missing mass was measured in the
neutron missing mass from the p~p~$\to$~p~$\pi^{+}$~X reaction at forward
angles. It was close to $\sigma(\Gamma) \approx$~2.2~MeV at
$\theta$=0.75$^{0}$
and T$_{p}$=1.805~GeV. At both incident energies, $T_{p}$=1520~MeV and
$T_{p}$=1805~MeV, the data were taken at
six angles from 0$^{0}$ (0.75$^{0}$) up to 17$^{0}$ in the lab. system.\\
\hspace*{4.mm}A simulation code was written in order to check the various
properties of the detection system. It reproduces the position and the
width of the neutron missing mass peak \cite{bor1}. The simulation code is
used to obtain the contributions of the peaks, by subtraction of the physical
smooth uncorrelated invariant masses (see later, section~III.B.).
\section{Description of the analysis.}
In the p~p~$\rightarrow$~p~p~X reaction, each invariant mass $M_{pX}$ studied,
has two different
kinematical solutions corresponding to different transferred momenta.
Fig.~2 exhibits the scatterplots of M$_{Xp_{i}}$ versus p$_{j}$
{\it (i$\neq$j)} at T$_{p}$=1520~MeV, $\theta=2^{0}$.
{\it i} and {\it j} denote the fast (f) and slow (s) detected protons.
We see clear cuts, created either by
physics or by the momenta limits of the spectrometer:
0.6$\le$pc$\le$1.4~GeV.
The data must be studied separately for the two branches, otherwise an
artificial peak structure will appear in the case of a simple addition.
All preliminary spectra, versus $M_{p_{f} X}$ or $M_{p_{s} X}$,
 will therefore be presented as belonging to either the
``upper branch'' or ``lower branch'', depending on whether they are above, or
below, the turnback limit in the scatterplots.
Here, $M_{p_{f} X}$ or $M_{p_{s} X}$
are the invariant masses obtained by using the momenta $p_{f}$ and
$p_{s}$ of the fast and slow emitted protons respectively. The two branches
were separated through software cuts on the momentum of the proton not included
in the invariant mass (see Fig.~2). These cuts create narrow
invariant mass regions with incomplete efficiencies. These narrow
ranges are therefore omitted in the spectra. In a few cases, the sum of the 
spectra from both branches is performed; the eliminated range due to incomplete
efficiency is delimited by vertical lines.
The missing masses vary from 0 up to 600~MeV at forward angles and for the
lowest incident energy. Such a range of results opens up the study of the
de-excitation of the baryonic resonances into N$\pi^{0}$ and N$\eta$
channels at forward angles (see Fig.~3). This range does not allow the study 
of the disintegration into heavier and still narrower mesons (and nucleons).
For increasing spectrometer angles, the resolution spoils whilst 
the maximum missing mass decreases. The first effect
results from the two terms: 
\begin{math}
\frac{\partial~M_{X}}{\partial~\theta_{3}}\hspace*{3.mm} 
\end{math}
\hspace*{-2.mm}and
\begin{math}
\frac{\partial~M_{X}}{\partial~\theta_{4}},
\end{math}
where $\theta_{3}$ and $\theta_{4}$ are the laboratory angles of both detected
protons. At a given angle, the resolution
improves for increasing missing mass; this property is due to the term
\begin{math}
dM_{X}=A/M_{X}
\end{math}\\
\hspace*{4.mm}At T$_{p}$=1520~MeV, the $\eta$ is no longer observable for
angles above 9$^{0}$. At this incident energy, the N$\pi^{0}$
de-excitation channel is observable at all six angles studied.\\
\hspace*{4.mm}At T$_{p}$=1805~MeV, the missing mass varies from 300~MeV up
to 750~MeV at forward angles, and varies from 0~MeV up to 500~MeV at
$\theta$=17$^{0}$. Therefore only the N$\eta$ de-excitation channel is
studied at forward angles, and N$\pi^{0}$ is studied at backward angles.\\
\hspace*{4.mm}The N$\pi^{0}$ (N$\eta$) de-excitation channel is
selected through the following
software cuts: M$_{X}\le$250~MeV (540$\le M_{X}\le$565~MeV). The range of
the invariant masses differs at both incident proton energies:
1470~MeV$\le$M$\le$1585~MeV at T$_{p}$=1520~MeV and 1560$\le$M$\le$1680~MeV
at T$_{p}$=1805~MeV.
Since these invariant masses are different, the data will be presented and
discussed separately in four different subsections.
\subsection{Comparison of momenta and missing masses with simulations}
The simulation code was written in the following way: the events were
generated using a flat random distribution of fast proton momenta, and of
angular
distributions: $\theta_{pf}$, $\theta_{ps}$, $\phi_{pf}$, and $\phi_{ps}$.
For angles other than forward ones, a variable distribution of $\theta_{pf}$
was introduced between the maximum and minimum possible angles. A Gaussian
distribution was introduced for the $\eta$ width. We take into account the
width of the
incident beam, the spectrometer resolution, and the detection resolution.
We do not introduce the description (mass and width) of any baryonic
resonance in the invariant M$_{p\pi}$ mass, nor do we introduce a specific
disintegration channel for this resonance. Our simulation therefore
describes the shape of uncorrelated events.
We expect to observe, if any, correlated p$\pi^{0}$ and correlated p$\eta$
events if these give rise to narrow structures in our
invariant mass range. We do not expect to be able to observe the 
broad PDG baryonic resonances since our mass range is too limited for that
and the resonances are mixed.
 In our relatively small invariant mass range the simulated
spectra must be continuous with a slowly varying slope. The measured spectra
can display discontinuous shapes,
since all reactions studied are exclusive.\\
\hspace*{4.mm}A comparison of detected and simulated proton momenta
 when the missing mass is selected by software cuts to the $\pi^{0}$,
is shown in Fig.~4. Inserts (a) and (b) show the momenta distributions
of the upper branches of M$_{\pi^{0}pf}$ and M$_{\pi^{0}ps}$. Full points
are data and empty points are the simulation results for T$_{p}$=1520~MeV and
$\theta$=2$^{0}$. Both lower branches are empty. Insert (c) shows the
missing mass obtained under the same conditions as in insert (a).
Fig.~5 shows similar histograms for T$_{p}$=1520~MeV again, but at
$\theta$=9$^{0}$.\\
\hspace*{4.mm}We observe the same global shape for data and simulation, 
at all angles, for the three variables shown in Fig.~2.
This shapes are smooth for simulated events and
quite dispersed, with non-negligible error bars, for data. 
\subsection{Subtraction of the simulated uncorrelated events}
This subtraction is illustrated in detail for the upper branch of
M$_{\pi^{0}ps}$ at T$_{p}$=1805~MeV, $\theta$=17$^{0}$. Insert (a) of
Fig.~6 shows the data (full points) and the corresponding normalized
simulated spectrum (empty points). Over the first half of the range, both
distributions nearly coincide, in the second half we observe many more
measured events than simulated events. The subtraction between the data and
simulated events is shown in insert (b). These events correspond to correlated
p$\pi^{0}$ from narrow N$^{*}$ or $\Delta^{+}$. These results will be
discussed in the forthcoming section IV.C.2.\\
\hspace*{4.mm}Another illustration is shown in Fig.~7 for the
 baryonic resonance de-excitation into the $\eta$n channel with
T$_{p}$=1520~MeV and $\theta=2^{0}$. We
observe, from Fig.~2 that the main statistics will appear from
the upper branch of M$_{Xps}$, whereas the corresponding lower branch will
give rise to only a few events (and will therefore be omitted). 
Fig.~7(a) shows the data (full circles) and the background from simulation
below the p$\eta$ peaks (empty circles). This last spectrum is used for
the uncorrelated event description (background). It is normalized
to the maximum possible number of events, but the points raised in this paper
would not be modified with a smaller selection of uncorrelated
events for the background choice. In Figt.~4(b) the difference between the
data and the simulated uncorrelated events is shown; it is decomposed into
three gaussians, each having the same width (see Table~I, first line).\\
\hspace*{4.mm}Fig.~8 shows, for three different kinematical
conditions, the data minus the normalized
simulations for the uncorrelated events, at T$_{p}$=1520~MeV, $\theta=2^{0}$.
The three inserts (a), (b), and (c) exhibit respectively the number of events
versus $M_{Xps}$ when $p_{f}\ge~p_{f}$limit,
versus $M_{Xpf}$ when $p_{s}\ge~p_{s}$limit, and versus $M_{Xpf}$ when
$p_{s}\le~p_{s}$limit. Here again, p~limit is the momentum value of the
turnback in the scatterplot of $M_{Xpi}$ versus $p_{j}$ where j$\neq$i.
Both Fig.~2 and Fig.~8 use the same data, but in the latter the N$\eta$ 
disintegration channel is selected through software cuts. The reduction in the
invariant mass range is particularly important for insert (b) of
Fig.~8. When applying the same selection at other angles, namely when we
observe M$_{Xpf}$ with p$_{s}$ larger than p$_{s}$limit, and select the
N$\eta$ disintegration channel, we obtain a small
increase in the statistics for increasing angles up to 9$^{0}$, followed by
a decrease. Given the weak statistics, this last selection of events will
 not be presented for this disintegration channel at this incident
proton energy.\\ 
\hspace*{4.mm}Table~II shows that the same mass values and
nearly the same widths are obtained
from a gaussian fit of the last insert (c) of Fig.~8. The
range of insert (b) is too small to allow to a peak to be extracted; the values
extracted from inserts (a) and (c) are consistent with the few existing data
points.
\subsection{Subtraction of  physical uncorrelated events and background
processes}
The ``background'' is constructed by a normalized addition of the total
invariant $M_{p\eta}$ mass in both parts, below and above the $\eta$ meson
mass in the missing mass spectrum. The software cuts chosen are
500$\le$M$_{X}\le$525~MeV and 575$\le$M$_{X}\le$600~MeV respectively for
both ranges. Fig.~9(a) shows the data points with the normalized ``background''
discussed before, whereas
insert (b) shows the subtracted spectrum, with three gaussian fits, as before.
In Table~I, the last line  shows the properties of the extracted peaks, which
all have the same mass and the same width as before when the
simulated ``background'' is used. The relative surface of the peaks, obtained
with help of both ``backgrounds'', moves a little, but has no influence on the
discussion which will be presented later. Such an agreement between masses and
widths, obtained with help of both choices for the ``background'', allows
us to keep the simulated ``background'' for the other data presented later on.
When the physical ``background'' is used, the ranges of background selected on
either side of the peak are at least 37.5~MeV away from the central peak. This
avoids including an eventual $\eta$ meson tail if it is present.
 One direct consequence of this is an irregular shape on
the high mass side of the background spectrum as shown in Fig.~9.
\section{Results and discussion}
Different final states will be studied successively: first, the final state
without any particular de-excitation channel, then the two final states with
the de-excitation into p$\pi^{0}$ (n$\pi^{+}$) and p$\eta$.
\subsection{The N$^{*}$ or $\Delta^{+}$ final state}
Partial results of the reactions studied were already shown in reference 
\cite{netw}.
As previously explained, different spectra must be considered that separate
upper and lower branches, and the invariant mass constructed with the slow
(fast) proton: M$_{psX}$ (M$_{pfX}$).\\
\begin{center}
{\it IV.A.1 The pp$\to$p$_{f}$p$_{s}$X reaction at
T$_{p}$=1520~MeV.}
\end{center}
\hspace*{6.mm}IV.A.1.(a)  Cross sections versus M$_{psX}$\\
\hspace*{4.mm}The cross sections of the upper branch of the
pp$\to$p$_{f}$p$_{s}$X reaction at T$_{p}$=1520~MeV, versus M$_{psX}$ are
shown in Fig.~10. The left-hand part shows the results at four forward
spectrometer angles: $\theta$=0$^{0}$, 2$^{0}$, 5$^{0}$, and 9$^{0}$. Clear
peaks are observed close to 1500 and 1535~MeV, and also, but less clear,
close to 1480~MeV (and possibly close to 1575~MeV). The maximum is close to
M=1535~MeV, the mass of the first S$_{11}$
baryonic resonance J$^{P}$~=~1/2$^{-}$, T~=~1/2, but here the observed width
$\Gamma_{t}\approx$25-40~MeV, is much smaller than that expected from the PDG
where the mean value is as large as 150~MeV. Indeed, PDG reports for
N$^{*}$(1535) a width varying from 57 up to 240$\pm$80~MeV. Although it is
rather difficult to draw a background below the M=1535~MeV peak, we estimate,
at $\theta$=0$^{0}$ and 2$^{0}$, that the maximum of this M$\approx$1535~MeV
peak amounts to $\approx$30$\%$ of the total excitation at this mass.
There is also, at this mass, room for contributions from other PDG baryonic 
resonances with smaller and larger masses,
and also contributions from N$\pi$ and N$\pi\pi$ incoherent phase space 
contributions. It therefore seems highly improbable that these peaks
are purely the PDG broad resonances, but are more likely to be simply
 parts of them. This point will be clarified later. 
We suggest that the reason for which the PDG data are so wide is due to
the lack of experimental precision in previous experiments.\\
\hspace*{4.mm}The right-hand part of Fig.~10 shows the effect of looking at
the same data but
using different binnings. The binning in insert (b), (3.2~MeV/channel),
is the same as the one used in the left-hand part of Fig.~10. It
corresponds to the experimental resolution. The binning of insert (d),
(32~MeV/channel), is close to the one used in $\pi$N$\to\pi$N and
$\pi$N$\to\pi\pi$N reactions \cite{man} \cite{cut}. A more recent
energy-dependent and single-energy partial-wave analysis of $\pi$N elastic
scattering data \cite{arn1}, presents the results binned in steps close to
20~MeV. The amplitudes of the photoproduction data were also presented with
a binning close to 18~MeV \cite{arn2}. Many other papers were published
from year to year by the same group \cite{arndt1}, testing the sensitivity
of different
reactions to the pion-nucleon coupling constant \cite{arn3}, or studying the
parameter of the multipole analysis in the region of the first $\Delta$
resonance \cite{arn4} or in the range of the N(1535) and N(1650) resonances
\cite{arn5}.\\
\hspace*{4.mm}
The cross sections of the lower branch of the
pp$\to$p$_{f}$p$_{s}$X reaction at T$_{p}$=1520~MeV, versus M$_{psX}$ are
shown in Fig.~11. The data at the six measured angles are shown. The
small range studied in this lower branch, does not allow a clear observation
of narrow structures. However, we observe a narrow peak at M$\approx$1576~MeV
at $\theta$=2$^{0}$,
and at M$\approx$1532~MeV at $\theta$=17$^{0}$. Narrow peaks at these two
masses were already observed in the upper branch, at the beginning of this
paragraph.\\
\hspace*{6.mm} IV.A.1.(b)  Cross sections versus M$_{pfX}$\\
\hspace*{4.mm}The differential cross sections of the upper branch of the
pp$\to$p$_{f}$p$_{s}$X reaction at T$_{p}$=1520~MeV, versus M$_{pfX}$, are
shown in Fig.~12. Several narrow structures are observed at nearly
all six spectrometer angles. There are masses close to 1550, 1567, and
1577~MeV at $\theta$=0$^{0}$, 1560~MeV at $\theta$=2$^{0}$, 1577~MeV at
$\theta$=5$^{0}$, 1534~MeV at $\theta$=13$^{0}$, and 1530~MeV at
$\theta$=17$^{0}$.\\
The differential cross sections of the lower branch of the
pp$\to$p$_{f}$p$_{s}$X reaction at T$_{p}$=1520~MeV, versus M$_{pfX}$, are
shown in Fig.~13. Here also, as was previously the case in the lower branch
of M$_{psX}$, it is not possible to observe any narrow structure
except at M$\approx$1567~MeV, and at only one angle, $\theta$=5$^{0}$.\\.
In short, the masses of narrow structures observed at
T$_{p}$=1520~MeV, in the pp$\to$p$_{f}$p$_{s}$X reaction, are close to:
1480, 1500, 1533, 1550, 1560, 1567, and 1576~MeV. The masses of the narrow
structures extracted from the cross sections of the
pp$\to$p$_{f}$p$_{s}$X reaction at T$_{p}$=1520~MeV, versus M$_{pfX}$, are
not as stable, when studied with respect to M$_{psX}$. They look like 
``sliding'' spectra. Such a behaviour could eventually be the result of
interferences between different resonances. An attempt was made to check
this assumption and to extract the relative ratio of amplitudes between
resonances of the same quantum numbers. The results of were not conclusive,
and will not be presented here. The small structures observed in these
spectra, versus M$_{pfX}$, are therefore not included in the summary
Table~VII.\\
\begin{center}
{\it IV.A.2 The pp$\to$p$_{f}$p$_{s}$X reaction at T$_{p}$=1805~MeV.}
\end{center}
\hspace*{4.mm}At this energy, only the upper branch cross-sections are
shown, since the lower branches, once again, cover a very small invariant
mass range ($\Delta$M$\approx$50~MeV) and  are populated by weak statistics.
The left-hand side  of Fig.~14 shows the cross sections of the upper part of the
pp$\to$p$_{f}$p$_{s}$X reaction versus M$_{psX}$. Narrow peaks are observed
close to the following masses, (where the question mark means a weaker
definition): M$\approx$(1621~?), 1635, and 1668~MeV at
$\theta$=0.75$^{0}$, M$\approx$(1620~?), 1637, 1669~MeV at $\theta$=3.7$^{0}$,
M$\approx$1667~MeV at $\theta$=6.7$^{0}$, M$\approx$1600~MeV at
$\theta$=9$^{0}$, and M$\approx$1657~MeV at $\theta$=13$^{0}$. However, the
last few peaks are small, and will therefore be omitted later on in Table VII.\\
\hspace*{4.mm}The right-hand side of Fig.~14 shows the cross
sections of
the upper part of the pp$\to$p$_{f}$p$_{s}$X reaction versus M$_{pfX}$. A
main maximum is observed in the four smallest angles at a mass close to
M$\approx$1638-1640~MeV. A small peak is observed close to
M$\approx$1660~MeV  in the
$\theta$=0.75$^{0}$, and $\theta$=6.7$^{0}$ data.\\
\hspace*{4.mm}In summary, the masses of narrow structures observed at
T$_{p}$=1805~MeV, in the pp$\to$p$_{f}$p$_{s}$X reaction, are close to:
(1600~?), (1621~?), 1639, 1657, 1660, and 1668~MeV. We observe again a shift,
$\Delta$M$\approx$35~MeV, between clear peaks, similar to the observations
made at small angles and shown in the left-hand side of Fig.~14. Several broad
baryonic resonances are reported by PDG in this mass range; however the widths
of the narrow observed peaks are much smaller than the widths reported by PDG.
\subsection{The n$\pi^{+}$ final state}
The reaction pp$\to$p$\pi^{+}$n was studied at the same angles and same
incident energies as the pp$\to$p$_{f}$p$_{s}$X reaction. The left-hand part
of Fig.~15 shows the cross sections versus the four smallest
spectrometer angles obtained at T$_{p}$=1520~MeV. The maximum, close to
M$\approx$1506~MeV, remains steady at the same mass, regardless of the
forward angle of observation, and a shoulder is observed at
M$_{n\pi^{+}}\approx$1540~MeV. The right-hand part of Fig.~15 shows
the cross-sections obtained at the two largest angles with maxima close to
1580~MeV and 1540~MeV respectively, at $\theta$=13$^{0}$ and $\theta$=17$^{0}$.
\subsection{The N$\pi^{0}$ de-excitation channel}
\begin{center}
{\it IV.C.1 The N$\pi^{0}$ de-excitation data from baryonic
invariant masses obtained with T$_{p}$=1520~MeV incident protons}
\end{center}
\hspace*{4.mm}Amongst the four different kinematical conditions shown in
Fig.~2, only two at forward angles contribute to the
M$_{p\pi^{0}}$ missing mass. They are the two upper branches for fast
and slow detected protons. The $\pi^{0}$ is not detected in the
missing mass for the two lower branches.
The results are obtained at the six measured angles: 0$^{0}$,
2$^{0}$, 5$^{0}$, 9$^{0}$, 13$^{0}$, and 17$^{0}$ and are shown in figures
16, 17, 18, 19, 20, and 21 respectively. In these figures the two inserts
(a) and (b) present the number of events of the upper branches,
versus M$_{Xps}$ and M$_{Xpf}$ respectively. Full
circles show the data, empty circles show the normalized simulated
uncorrelated spectra,
considered to be the `background'', and full triangles show the difference
between them. In some figures, gaussian peaks are extracted from these
last spectra. The
quantitative values of these gaussian fits are reported in Table~III. 
 Although
the number of standard deviations found (S.D.) is large, the definition of the
peaks depends strongly on the uncorrelated event distributions,
and are therefore poorly defined at $\theta$=2$^{0}$ and
$\theta$=9$^{0}$. At $\theta$=13$^{0}$ (Fig.~20) and
$\theta$=17$^{0}$ (Fig.~21) three peaks
are extracted. The extracted masses of these peaks, at both angles, are about
the same (see Table~III). These patterns may indicate an oscillatory shape.
Table~III shows a more or less constant mass interval
between several peaks ($\Delta M\approx$11~MeV), since the masses of the
extracted peaks are close to M$\approx$1479, 1490, 1513, 1520, and 1532~MeV.
In this Table R shows the ratio of the peak surfaces (higher mass peak surface
divided by lower mass peak surface) where the two peaks
are observed under the same kinematical conditions.\\
\hspace*{4.mm}At non-forward angles, namely at $\theta$=13$^{0}$ and
$\theta$=17$^{0}$, there are events from the lower branches (see Fig.~2).
Fig.~22 shows the number of events versus M$_{\pi^{0}pf}$ at
$\theta$=13$^{0}$ (17$^{0}$) in insert (a) ((b)) respectively, at
T$_{p}$=1520~MeV.  The statistics are very low when shown as a function of
M$_{ps\pi^{0}}$, and the corresponding data were therefore not studied.
In the spectra showing the event population versus M$_{\pi^{0}pf}$, two peaks
are extracted at the same masses as for the upper branch (Fig.~13), namely
at M$\approx$1520~MeV and M$\approx$1531~MeV (see Table III). \\
\begin{center}
{\it IV.C.2 The N$\pi^{0}$ de-excitation data from baryonic
invariant masses obtained with T$_{p}$=1805~MeV incident protons}
\end{center}
The corresponding histograms are empty at forward angles and very few
events were accumulated at $\theta$=9$^{0}$. The data were studied at the
two largest angles: $\theta$=13$^{0}$ and $\theta$=17$^{0}$. At these
angles the number of events is no longer comparable to the forward angle
population since the lower branches are not negligible with respect to the
upper branches.\\
\hspace*{4.mm}Fig.~23 shows the events at $\theta$=13$^{0}$ of the upper
branch versus M$_{ps\pi^{0}}$: data (full
points), simulation (empty points) in insert (a) and difference between
the two, in full triangles in insert (b). There is a hole close to
M$_{ps\pi^{0}}$=1600~MeV which arises from the region where both p$_{f}$ and
p$_{s}$ detected proton momenta are similar.
In order to avoid such invariant mass ranges with lost events, the following
software cuts were applied: all events were suppressed (M$\approx$1600~MeV),
when the difference between both momenta was lower than 50~MeV/c.
The mass range 1595$\le$M$\le$1605~MeV was therefore removed from Fig.~23.
The quantitative information is given in Table~IV.
Fig.~24 shows the corresponding events from the upper branch, versus
M$_{pf\pi^{0}}$. Again, there is a hole in
the data close to M$_{p\pi^{0}}$=1611~MeV, which correponds to the region
where both proton momenta are equal. However, the simulated results show that
the hole is a real physical effect. The data show two
peaks described by two gaussians centered at 
following masses (widths) M=1605.7 ($\sigma(\Gamma)$=3.4)~MeV and M=1622
($\sigma(\Gamma)$=5.4)~MeV, rather than a single gaussian at M=1616
($\sigma(\Gamma)$=8.7)~MeV. However, since a (small) ambiguity due to the lost
events remains, these results are removed from
the general summary and from Tables IV and VII.\\
\hspace*{4.mm} A careful detailed analysis was performed at the largest
spectrometer angle ($\theta$=17$^{0}$). Figures~6 and 25 show the number
of events of the upper branches versus the  invariant masses M$_{ps\pi^{0}}$
and M$_{pf\pi^{0}}$,
respectively, at $\theta$=17$^{0}$, and T$_{p}$=1805~MeV. Fig.~26
 shows the sum of the data shown in Fig.~6(b) and
Fig.~25(b) (i.e. the data minus the simulation of both upper branches). 
The lower branches contain few events, and Table~IV shows that when all
upper and lower branches are added (last line of Table~IV), the peaks are
extracted at the same mass values which were observed before the addition
of the lower branches. A check was performed to ascertain that the dip
at M$_{p\pi^{0}}\approx$1607~MeV was not produced by an instrumental cut or
created by a software cut.
With this end in view, software cuts were applied to remove the range
1600$\le$M$_{p\pi^{0}}$$\le$1610~MeV in the upper branch data,
and the effect on the scatterplot
p$_{f}$ versus p$_{s}$ was studied. We observe, as shown in Fig.~27, that
with this software cut there is an overall reduction in the statistics over
the whole scatterplot. This means that the discussed range is not connected
to a particular range of momenta. Therefore, the dip is from a physical source
as are the two peaks. Quantitative  information is given in Table~IV.\\  
\hspace*{4.mm}From these measurements at T$_{p}$=1805~MeV, five peaks close
to the following masses are extracted: 1534~MeV, 1558~MeV, 1582~MeV,
1601~MeV, and 1622~MeV. The three first masses are observed only once,
but the last two masses are observed six and four times respectively.
The shift between these
five masses is more or less constant: $\Delta$M$\approx$22~MeV. 
\subsection{The N$\eta$ de-excitation channel}
\begin{center}
{\it IV.D.1 The N$\eta$ de-excitation data from baryonic
invariant masses obtained with T$_{p}$=1520~MeV incident protons}
\end{center}
Fig.~28 shows the missing mass M$_{X}$ at T$_{p}$=1520~MeV,
$\theta$=0$^{0}$, for the four kinematical situations. Inserts (a),
(b), (c), and (d) correspond to the upper part of p$_{s}$, the upper
part of p$_{f}$, the lower part of p$_{s}$, and the lower part of
p$_{f}$ respectively. The $\eta$ is clearly observed in inserts (b) and
(c). The data corresponding to the conditions of inserts (a) and (d) will
not be studied. The $\eta$ is selected by applying
the following software cut: 540$\le$M$_{X}\le$565~MeV.\\

\hspace*{6.mm}{\it IV.D.1(a)  Study of p$_{pf}\ge$p$_{pf}$limit data.}

\hspace*{0.5cm}The results for the smallest angles, where the
$\eta$~meson is observed
in our experimental conditions are shown. Figs.~29, 7, 30, and 31
show the number of events of the upper branch of the invariant mass
M$_{ps\eta}$ for $\theta$=0$^{0}$, 2$^{0}$, 5$^{0}$, and 9$^{0}$
respectively.
The values of the extracted peaks are given in Table~V. The peak at
M$\approx$1564~MeV is not well defined, and is only tentatively
extracted at
$\theta$=0$^{0}$ and 5$^{0}$ by analogy with its behaviour at
$\theta$=2$^{0}$ and 9$^{0}$ where some points remain on the high tail side.
Three peaks are extracted at M$_{p\eta}\approx$1502~MeV, 1519~MeV, and
tentatively at 1564~MeV.
The R values give the relative angular variation of the surface of the
three extracted peaks to the surface of the M$\approx$1564~MeV peak. They
are listed in Table~V and are drawn in Fig.~32.
This figure shows the relative peak
intensities: M$\approx$1505~MeV peak over M$\approx$1564~MeV peak (full
circles), and M$\approx$1517~MeV peak over M$\approx$1564~MeV peak (full
triangles).
They must be considered as
being tentative, since the surfaces are somewhat imprecise. We observe an
increase of the second ratio with increasing angles. It is not possible to
use this result to discuss the relative spins of both states
M$\approx$1517~MeV and M$\approx$1564~MeV. Indeed the transferred momenta vary
differently for these different final states for increasing angles. The two
other ratios are consistent with a constant value.\\
\hspace*{4.mm}Three masses are extracted from
our spectra, at M$\approx$1502, 1520, and 1565~MeV. The second mass is
the same as the one extracted with a software selection to N$\pi^{0}$.
The width here is larger, $\sigma(\Gamma)\approx$8~MeV instead of
$\sigma(\Gamma)\approx$4~MeV.
There is a peak at the same mass reported by PDG \cite{pdg}, namely the
D$_{13}$, but with a total width $\Gamma\approx$120~MeV. No peak was extracted
at M$\approx$1502~MeV from the N$\pi^{0}$ de-excitation data. There is no broad
resonance reported at either M$\approx$1502~MeV, nor at M$\approx$1564~MeV.\\

\hspace*{6.mm}{\it IV.D.1(b)  Study of p$_{ps}\le$p$_{ps}$limit data.}

Figs.~33, 34, 35, and 36 show
the number of events versus the lower branch of M$_{pf\eta}$ for angles
$\theta$=0$^{0}$, 2$^{0}$, 5$^{0}$, and 9$^{0}$ respectively, at
T$_{p}$=1520~MeV. For this 
kinematical configuration, $\theta$=9$^{0}$ is again the maximum possible
spectrometer angle. The quantitative information is given in Table~V.
Three peaks are extracted, a peak at M$_{pf\eta}\approx$1553~MeV, seen only
once at $\theta$=9$^{0}$ and therefore less well defined than the two others,
and two peaks close to M$_{pf\eta}\approx$1562~MeV and 1579~MeV. The
mass of the
first of these two peaks is very close to a mass observed previously
in different data, namely the upper part of M$_{ps\eta}$, since
the mass difference is only $\Delta$M$\approx$2~MeV. There is no broad
PDG baryon with any of these three masses. Column R in Table~V, and Fig.~32
show the relative peak intensities with respect to the M$\approx$1564~MeV peak.
The angular variation of this ratio is flat, and may be
considered as being consistent with a value of 1. 

\begin{center}
{\it IV.D.2 The N$\eta$ de-excitation data from baryonic
invariant masses obtained with T$_{p}$=1805~MeV incident protons}
\end{center}
Fig.~37 shows in the missing mass at T$_{p}$=1805~MeV, $\theta$=0.75$^{0}$,
clear M$_{\eta}$ peaks, in inserts
(a) and (c), superposed on large backgrounds. Inserts (b) and (d) show
the corresponding invariant masses, namely the upper branch of M$_{Xpf}$ in
insert (a) and the upper branch of M$_{Xps}$ in insert (b). Here the full
points correspond to all events and the empty points correspond to events
selected by software cuts to choose p$\eta$ disintegrations.
In the missing mass at T$_{p}$=1805~MeV, $\theta$=0.75$^{0}$,
clear M$_{\eta}$ peaks are observed, superposed on large backgrounds.
The statistics of the lower branch are small and no $\eta$ meson
peak is clearly observable in the missing mass spectra. This is illustrated in
Fig.~38. In this figure, insert (a) shows the addition of both upper
branch data, to be compared to insert (b) which shows the addition of both
lower branch data. Data between both vertical lines are removed since they
correspond to the transition region between M$_{pf\eta}$ and M$_{ps\eta}$
and where some events are lost as the result of the software cuts applied 
between upper
and lower branches. Insert (c) shows the total spectrum of all branches.\\
\hspace*{4.mm}Insert (c) of Fig.~38 shows a peak superposed on a
background. This background  corresponds to uncorrelated events and is
studied on Fig.~39 with help of our simulation code. Insert (a) of
Fig.~28 shows the renormalized simulated events at T$_{p}$=1805~MeV,
$\theta$=0.75$^{0}$, from the upper branch of M$_{pf\eta}$ (empty marks) and
from the upper branch of M$_{ps\eta}$ (full marks). These simulated data
are fitted by two
straight lines. Insert (b) shows the same results for $\theta$=3.7$^{0}$.
The normalized lines fitting the ``background'' are reported in insert (c)
of Fig.~38.\\
\hspace*{4.mm}Fig.~40 shows the data for $\theta$=3.7$^{0}$ presented in the
same way as in Fig.~38 (for $\theta$=0.75$^{0}$). The simulation and
data for $\theta$=6.7$^{0}$ are shown in Fig.~41 and
Fig.~42.\\
\hspace*{4.mm}Fig.~43 shows the spectra extracted from the
pp$\to$p$_{f}$p$_{s}\eta$ reaction at T$_{p}$=1805~MeV at the three
forward angles $\theta$=0.75$^{0}$, 3.7$^{0}$, and 6.7$^{0}$
corresponding to inserts (a), (b), and (c) respectively. These data came
from the subtraction of total events minus uncorrelated simulated events
shown in Figs.~38, 40, and 42 for  angles
$\theta$=0.75$^{0}$, 3.7$^{0}$, and 6.7$^{0}$ respectively.
Two peaks at M$\approx$1642~MeV and M$\approx$1658.5~MeV are extracted at
all three angles. The quantitative
information is given in Table~VI. The ratio R, shown in Fig.~44,
gives the angular distribution of the relative excitation between the
M$\approx$1642~MeV and the M$\approx$1658.5~MeV peaks. The error bars are
arbitrarily put to $\delta$R/R=0.15. As previously explained, the
different transferred momenta variation, as a function of the angle,
prevents drawing any conclusion on the different spins of these two states.\\
\hspace*{4.mm}Although the missing mass range is 
not larger than 70~MeV, the observed shape is not the consequence of a low
acceptance due to cuts applied at
both edges of the range. These cuts can be clearly
seen on both sides of the spectra in Figs.~38(c), 40(c), and 42(c),
 and affect 4 or 5 channels only on each side. 
\section{General discussion or Comparison between both de-excitation
branches.}
Several N$^{*}$ and $\Delta$'s are reported in \cite{pdg} to
exist in the mass range 1440$\le$M$\le$1700~MeV. They all appear with four
stars in the N$\pi$ de-excitation mode. Two baryonic resonances are
reported to have four stars in the N$\eta$ desexcitation mode: the N(1535)
and N(1650) both  S$_{11}$. In this channel the observed structures cannot
come from any $\Delta$ resonance.\\
\hspace*{4.mm}All our structures have smaller widths than those reported in
\cite{pdg} for the baryonic resonances. They are also weakly excited 
compared to the excitation of the PDG resonances. It is therefore tempting to
associate our narrow structures to the PDG broad ones, providing the latter
have a ``fine structure''.\\
\hspace*{4.mm}By analogy with the nuclear giant resonances, we can argue
that the baryonic resonances are so broad, that their lifetime is smaller
than 10$^{-23}$s. Their de-excitation takes place very quickly, without
time for internal reorganizations. Therefore the information on their
microscopic structure is conserved. The narrow observed peaks may be
signatures of these internal structures.\\
\hspace*{4.mm}We observe that interferences between two broad classical
baryonic resonances never produce peaks in the middle region between
them, but they do produce shifts, usually by $\pm$10 ($\pm$20)~MeV of the
maximum of the considered resonance.
Therefore, the many narrow observed peaks cannot be the result from
interferences.\\
\hspace*{4.mm}Table~VII shows the quantitative information for the narrow
peaks extracted. Some peaks, being too poorly defined, are
omitted. The overall data are plotted in Fig.~45, where the channel is defined
in Table~VII. The dashed areas show the range of study allowed by the
physics and the spectrometer momenta acceptance. All broad baryonic PDG
resonances have an important N$\pi$ de-excitation channel. We make the
assumption that the same is true for the narrow baryonic resonances, even if
the spectrometer limits prevent their detection.
We observe in Fig.~45 that the many masses of narrow peaks M$\pm\Delta$M,
where $\Delta$M=3~MeV, can be brought together into 13 masses shown in the
right-hand part of the figure. The full (empty) marks indicate the well
(less well) defined peaks. \\
\hspace*{4.mm} The mean values of the extracted masses are close to: 1479,
1505, 1517, 1533, 1542, 1554, 1564, 1577, 1601,
1622, 1639, 1659, and 1669~MeV. Most observations have been made several
times at the same mass, as can be seen in Fig.~45. However, they have
different disintegration modes. As previously discussed, the
pp$\to$pp$\pi^{0}$ and pp$\to$p$\pi^{+}$n reactions are different from
the dynamical point of view since in the first one the $\pi^{0}$ is not
detected
and has a small momentum, whereas in the second reaction the $\pi^{+}$ is
detected and has therefore a higher momentum (p$_{\pi^{+}}\ge$600~MeV/c).\\
\hspace*{4.mm}Some masses are seen in both N$\pi$ and N$\eta$ channels,
some others are
seen in the N$\pi$ channel and not in the N$\eta$ channel. One mass
M=1564~MeV is seen in the N$\eta$ channel and is not seen
in the N$\pi$ channel, although the experimental acceptance allows such an
observation. Fig.~46 shows the same results. The
left-hand part shows the masses disintegrating through the N$\eta$ channel,
and the right-hand part the masses disintegrating through the N$\pi$ channel.
The broad dark lines correspond to broad PDG resonances. The vertical
lines are dashed when the disintegration mode is not observed, but is
likely to exist.
\subsection{Masses disintegrating into N$\pi$ and N$\eta$}
Two groups of masses disintegrate through N$\pi$ and N$\eta$ modes. The first
one corresponds to the following masses: 1505~MeV, 1517~MeV, 1554~MeV,
1564~MeV, and 1577~MeV. We tentatively associate these five masses to the
broad PDG S$_{11}$(1535) resonance. Indeed, the mean value of our five masses,
$\bar{M}\approx$1543~MeV, is close to the PDG mean mass value: 1535~MeV. The
gap between the extreme masses: 72~MeV, fills half of the estimated PDG width:
150~MeV. That several narrow-resonance baryons have an N$\eta$
disintegration channel may, eventually, agree with a calculation of the total
the pp$\to$pp$\eta$
cross section \cite{ceci}. Ceci {\it et al.} \cite{ceci} ``emphasize the fact
that a single resonance model, using only N(1535) drastically fails to
describe the experimental data. The next S$_{11}$ resonance N(1650) must be
included, and the introduction of the third controversial S$_{11}$ N(2090)
represents a further improvement''.\\
\hspace*{4.mm}The second group of masses which disintegrate through N$\pi$ and
N$\eta$ modes, holds two masses, namely M=1639~MeV and M=1659~MeV. Here
$\bar{M}\approx$1649~MeV is very close to the PDG mass value of the
second S$_{11}$ resonance: 1650~MeV.
The gap between both masses: 20~MeV is much smaller than the PDG
estimated width: 150~MeV.
\subsection{Masses having no N$\eta$ disintegrating channel}
Here again, a more or less large gap allows us to tentatively divide the
masses of these narrow resonances into two groups. The first one
corresponds to the following masses: 1479~MeV, 1533~MeV, and 1542~MeV.
Here $\bar{M}\approx$1518~MeV and is very close to the PDG mean mass value
of the D$_{13}$(1520) resonance. The gap between extreme masses: 63~MeV
corresponds, as it was in the case of S$_{11}$(1535), to half of the estimated
width of the broad PDG resonance.\\
\hspace*{4.mm}The second group of masses which do not have the N$\eta$
disintegration mode corresponds to the following masses: M=1601~MeV and
1622~MeV. Here $\bar{M}\approx$1611~MeV, is close to the PDG mean mass
value of the $\Delta$ P$_{33}$(1600) resonance: 1600~MeV.\\
\hspace*{4.mm}It appears therefore that all our masses of narrow baryonic
resonances can be associated with PDG resonances, provided the latter
can be split into several narrow resonances.
Only one, at M=1669~MeV, could be a part of the D$_{15}$(1675) PDG
resonance. However, it could also be associated with several hypothetical
heavier narrow resonances, which are not observed since their masses lie
outside our mass acceptance.\\  
\subsection{Discussion concerning the branching ratios}
Since we observe narrow baryonic resonances which disintegrate into N$\pi$
and N$\eta$, it is tempting to compare the ratios of the number of events 
between both channels.\\
\hspace*{4.mm}The comparison between the p$\eta$ final state data (channels
$\delta$ to $\nu$ in Fig.~45) and data obtained without final state
definition (channels a to k in Fig.~45) is meaningless.\\
\hspace*{4.mm}The comparison between the  p$\eta$ final state data (channels
$\delta$ to $\nu$ in Fig.~45) and
n$\pi^{+}$ final state data (channels
l to q in Fig.~45) is again meaningless. Indeed, although we have
data at the same angles ($\theta$=0$^{0}$ and 2$^{0}$) from the same branch,
namely the upper branch, for both reactions, the final states are totally
different, since in one case the $\pi^{+}$ is detected and the neutron is
slow, and in the other case, the proton is detected and the $\eta$ is slow.\\
\hspace*{4.mm}The comparison between the p$\eta$ final state data (channels
$\delta$ to $\nu$ in Fig.~45) and the p$\pi^{0}$ final state data
(channels r to $\eta$ in Fig.~45) is more promising. Indeed we have, in both
channels,
data showing a narrow peak at M=1517~MeV at the same incident proton energy
(T$_{p}$=1520~MeV), from the same branch (the upper branch), and using
the same slow proton for both reactions. However the $\eta$ is selected
at small angles (from $\theta$=0$^{0}$ to 2$^{0}$, channels $\delta$ to
$\epsilon$ in Fig.~45), when the $\pi^{0}$ is selected at large
angles (from $\theta$=13$^{0}$ to 17$^{0}$, channels t to x in Fig.~45).
These limitations result from the spectrometer momenta limits.\\
\hspace*{4.mm}We conclude that we are unable to give relative branching
ratios. \\
\hspace*{4.mm}It is worthwhile to point out that a large ``background'' of
2 pions exist, whose branching ratios cannot be compared with the N$\pi$ or
N$\eta$ branching ratios, since our mass (and angular) range is too small.
\section{Comparison with other results}
Occasionally data, published in order to study different problems,
 display narrow discontinuities which are not discussed by
the authors. These narrow discontinuities are sometimes, but not always,
well defined statistically. The authors of such results did not take into
account the possibility
to associate the discontinuities of their spectra with possible narrow
baryonic low mass structures. We present and discuss some such results and 
limit our discussion to the mass range
1.47$\le$M$\le$1.68~GeV, which is the range studied in this work.
\subsection{The p($\alpha,\alpha'$)X reaction}
A precise spectra of the p($\alpha,\alpha'$)X reaction was obtained twelve
years ago at SPES4 (Saturne) in order to study the radial excitation of the
nucleon in the P$_{11}$(1440~MeV) Roper resonance. A spectrum measured at
T$_{\alpha}$=4.2~GeV, $\theta=0.8^{0}$ \cite{morsch1} was defined in the
baryonic missing mass range: 1030$\le$M$\le$1490~MeV. A peak is extracted
at M=1478~MeV, very close to M=1479~MeV where a peak was extracted from our
data. Another spectrum was measured at T$_{\alpha}$=4.2~GeV, $\theta=2^{0}$
\cite{morsch2} which extended up to M=1588~MeV. Fig.~47 shows the part of
this spectrum for masses larger than 1470~MeV. The statistical errors could
not be larger than a factor of 2 from the those extracted using the given
counts \cite{morsch3}. The error bars are therefore multiplied by a
factor of 2. The comparison of the masses of the narrow structures extracted
from the experiments performed with the SPES4 and SPES3 beam lines is
shown in Fig.~48 and Table VIII. The positions of the structures,
observed in both experiments,
agree to a high extent. This agreement is obtained from published data,
originally obtained from experiments with different objectives, carried out
by different physicists and using different set-ups, beams, and reactions.
\subsection{Total cross-sections of $\pi$N reactions}
Rather old total cross-sections of $\pi$N reactions were reported in a CERN
compilation \cite{cern}. Whereas neither $\sigma_{T}(\pi^{-}$p) nor
$\sigma_{T}$ ($\pi^{+}$p) \cite{cart} display any narrow peaks, the data
from $\sigma_{T}$ ($\pi$p$\to$n$\pi^{0}$)  
\cite{cern} \cite{bulos}, show small narrow and not very precise peaks 
 close to M=1500~MeV [1505], 1635~MeV [1639], and 1660~MeV [1659].
Here the numbers between square brackets are the masses extracted in this work.
\subsection{$\eta$ meson photoproduction on the proton}
The total cross section of the $\eta$ meson photoproduction on hydrogen was
measured near threshold at ELSA (Bonn) \cite{pri}. The very small incident
photon energy range
prevents the possibility of observing baryonic structures.
Total and differential cross sections for the pp$\to$pp$\eta$ reaction were
studied at the internal beam facility at COSY \cite{mosk}. In this case the
total range of M$_{p\eta}$ is also very restricted.
The same reaction was also studied at the MAMI accelerator in Mainz
\cite{kru} \cite{kru1}.
Total cross-section data are plotted in Fig.~49. Three small and narrow
peaks (total width $\approx$ a few MeV) can be
extracted, even if the definition of the second one (a2) is better than are
the definitions of the two other peaks. The masses of these three peaks
can be compared to the first three masses in our data,
above the threshold of $\eta$ meson photoproduction off the proton. These
masses fit accurately with the masses extracted from the SPES3 data, as
can be seen in Table IX.\\
\hspace*{4.mm}The differential cross section of $\eta$ photoproduction on
the proton was
also measured at GRAAL (Grenoble) \cite{rena}. Here again, the resolution
(and binning) is
not appropriate for the study of narrow peaks.
\subsection{Near threshold electroproduction of $\eta$ meson on the proton}
 The differential
cross-section of the p(e,e'p)$\eta$ reaction was studied at JLab in Hall C
\cite{arms}. The $\eta$ electroproduction was also studied at JLab with the
CLAS spectrometer \cite{muel}. The integrated cross-sections show small
peaks
in the vicinity of 1561~MeV, 1582~MeV and 1621~MeV. Here again, the binning
of 20~MeV prevents a better observation of any eventual narrow peak.
However, the masses are close to the those extracted from the SPES3 data.
\subsection{The $\gamma$p$\to$n$\pi^{+}$ reaction}
Differential cross sections for the $\gamma$p$\to$n$\pi^{+}$ reaction were
measured at the Bonn 2.5~GeV electron synchrotron \cite{dann}. The
bremsstrahlung beam allows measurements over a wide incident energy range,
typically 0.31$\le$T$_{\gamma}\le$2~GeV at six angles between
$\theta_{\pi}$=180$^{0}$ and 95$^{0}$. Additionnal data, in a more limited
 photon energy range were taken at six outgoing pion angles between
 $\theta$=85$^{0}$
and 35$^{0}$ \cite{fische}. Fig.~50 shows two small peaks extracted
at $\theta$=65$^{0}$ at M=1.638~GeV ($\sigma(\Gamma)$=10~MeV), and at M=1.689~GeV
($\sigma(\Gamma)$=8~MeV). The first of these masses is very close to one of our
narrow peak masses (M=1639~MeV) and the second one is a little higher than
our experimental acceptance.\\
\hspace*{4.mm}The differential cross section of the reaction
$\gamma$p$\to\pi^{+}$n was measured with PHOENICS at ELSA (Bonn) over a large
range of photon energy: 220$\le$E$_{\gamma}\le$900~MeV \cite{buch}. However,
the wide binning implies that any narrow peak would be washed out, and so,
be unobservable.\\
\subsection{Meson photoproduction on the nucleon}
Many cross sections were measured on the proton and neutron, with one or more
pions produced. A review article lists the corresponding results
\cite{krusche}. Several other recent experiments were performed with a
binning which does not allow to distinguish narrow peaks.
In the mass range studied in this work, many recent results exist
which have been measured at the CEBAF Large Acceptance spectrometer and at
other accelerators at MAMI
(Mainz), ELSA (Bonn), GRAAL (Grenoble) \cite{assa} to name a few.
The resolution (and the binning) of these data are too weak, preventing the
possibility of observing
narrow structures such as those shown previously. This is typically the case
for the $\gamma$p$\to$p$\pi^{0}\pi^{0}$ total cross section measured at GRAAL.\\
\hspace*{4.mm}The cross section for the reaction ep$\to$e'p$\pi^{+}\pi^{-}$
was measured at JLab (CLAS) in the resonance region 1.4$\le$W$\le$2.1~GeV
\cite{rip}. The authors concluded on the presence of resonant structures
that were not visible in
previous experiments. The $\eta$ meson electroproduction cross section was
also measured at JLab (CLAS) with the center of mass total energy
1.487$\le$W$\le$1.635~GeV
\cite{tho}. The authors concluded on some indication of a $Q^{2}$ dependence
of the width of the N(1535) S$_{11}$  resonance, although the data are not
conclusive. They also attributed to an interference between S and P waves,
the new structure they observed at W$\approx$1.65~GeV.\\
\hspace*{4.mm}Photon and $\pi^{0}$ electroproduction from hydrogen were
studied at JLab Hall A \cite{lave}. The excitation curves, presented with a
binning of 20~MeV, show an oscillatory pattern of possible peaks with
widths that would be much smaller than those ($\Gamma~\approx$200~MeV) of
classical baryonic resonances.\\
\hspace*{4.mm}Total photoabsorption cross sections were also measured at
MAMI (Mainz) \cite{mario} in the photon energy range
200$\le$E$_{\gamma}\le$800~MeV, for several target nuclei. In this last
data set, the binning is smaller $\approx$8~MeV, in the region of M=1.5~GeV,
but the total cross section is not the best channel for observing
narrow exotic, and therefore small, effects.
\section{Comparison with models of baryonic resonances}
The classical baryonic spectrum, in the mass region studied here, was
analyzed through a partial-wave analysis of pion-nucleon elastic scattering
data \cite{arn1} and partial-wave analyses of single-pion photoproduction
data \cite{arn2} \cite{arndt1}. In both cases, the data were obtained with
binnings close to 20~MeV.\\
\hspace*{4.mm}The total and differential cross-sections of the baryonic
production in nucleon-nucleon reactions, were calculated in the one-boson
exchange model \cite{hub}. The authors concluded that their model, after
adjustement to the elastic nucleon-nucleon scattering, agree with the
experiment.\\
\hspace*{4.mm}Many models of baryonic resonances were proposed (see ref.
\cite{cap}). The quark models are restricted to the assumption of
$|q^{3}>$ wave functions, without considering additionnal $q\bar{q}$ pairs
or gluonic
contributions \cite{cap}, \cite{cap1}, \cite{isgu}. These calculations are
often related to the search for ``missing baryons'' which possibly couple
weakly to the N$\pi$ channel \cite{cap2}. The low-lying baryon spectrum was
calculated within the chiral constituent quark model \cite{gar}. The
light-baryon spectrum was calculated within a relativistic quark model with
instanton-induced quark forces \cite{lori}. On the basis of the
three-particle Bethe-Salpeter equation, a good description of the overall
baryonic mass spectrum up to the highest spin states was obtained
\cite{metsch}.\\
\hspace*{4.mm}The properties of baryon resonances in the mass range studied
in this paper,
were calculated \cite{dytm} using $\pi$N data and a multichannel unitary
model. The authors found ``results similar to previous analyses for strongly
excited states, but the results can vary considerably when the states are
weak''.
They emphasize ``that the full width of the S$_{11}$(1535) varies largely,
due to the close proximity of the resonance pole to the $\eta$N threshold''.
It is worthwile to note that the widths found in the calculation of
\cite{arndty} are consistent, for the S$_{11}$(1535), with our
``experimental total'' width obtained by combining the shift between the
extreme masses: 1505~MeV and 1577~MeV, and the width of each narrow structure.
Indeed, the calculated width in \cite{dytm} for $\eta$N is
$\Gamma$=66$\pm$13~MeV, and the calculated width for $\pi$N is
$\Gamma$=77$\pm$17~MeV.\\
\hspace*{4.mm}Several recent works investigate the baryon spectroscopy in
lattice QCD \cite{lein}, \cite{chiu}. 
The discussion of these results is outside the scope of the present work.  
\section{Conclusion}
Using the SPES3 spectrometer and detection system at Saturne, the
pp$\to$p$\pi^{+}$X
and pp$\to$ppX reactions were studied at two incident energies:
T$_{p}$=1520~MeV and 1805~MeV. With the help of software cuts on the missing
mass spectra,
the following final states were selected: p$\pi^{0}$, n$\pi^{+}$,
and p$\eta$. The N$\pi\pi$ final state contributed to the physical
background. Due to the good resolution and reasonable statistics, we were
able to observe in the range 1.47$\le$M$\le$1.68~GeV, a large number of
narrow and
well separated peaks. In some cases the peak to background ratio is not
large, therefore a peak is only considered as a narrow baryonic resonance   
when several peaks are extracted at the same mass ($\pm$3~MeV), and in  
different experimental data sets. The widths of these peaks are low as
compared to the widths reported by the PDG.\\
\hspace*{4.mm}We have separated the narrow states observed into two classes:
those where a
de-excitation into a N$\eta$ final state was observed and those where this
was not the case. When combining the masses of narrow states seen
in both N$\eta$ and N$\pi$  final states, we tentatively identify them as
being ``fine structures'' of N$^{*}$(1535)S$_{11}$ and N$^{*}$(1650)S$_{11}$.
In the same way, the masses not observed in the N$\eta$ final state are
tentatively identified as being ``fine structures'' of N$^{*}$(1520) and
$\Delta$(1600)P$_{33}$. Such assumptions give, for each case, a mean mass
that is in fairly good agreement with the mass of the broad PDG baryonic
resonance. Such a description could also justify the different mean masses
sometimes observed when different reactions are used. For example the mass
of the N(1535)S$_{11}$ was found at M=1549$\pm$2~MeV from the
$\pi^{-}$p$\to\eta$p reaction \cite{abaev} and at M=1525$\pm$10~MeV from the
$\gamma$N$\to\pi$N reaction \cite{arn2}. Also, the calculated partial widths
are generally too small to account for the experimental values of broad PDG
baryons \cite{metsch1}.\\
\hspace*{4.mm}By analogy with nuclear physics, we suggest
that the previously broad PDG baryonic resonances, are in fact collective
states of several weakly excited and narrow resonances. These 
resonances, can be single-particles or quasi-particles (from constituent
quark) states, with quark structures  more complicated than $|q>^{3}$.
We suggest that the reason for which these narrow weakly excited peaks were
not observed till now, is due to the lack of experimental precision of
previous experiments.\\
\hspace*{4.mm}
It is clear that these observations - if confirmed by other
experiments - would be a milestone in  hadron spectroscopy, and would impose a
big challenge to hadron theorists. The hadron structure is more rich than
has been often though up till now. Its study
requires non-pertubative methods in the mass range studied in this paper. It
 remains to explain the small widths observed. One possibility may be that
there exists some ``quark tunnelling'' from one quark-cluster to another. 
It is then possible that new theoretical tools will have to be developed.
Indeed, a complete description of hadrons should incorporate valence quarks,
sea quarks, and gluonic degrees of freedom.\\

We thank Pr. H. Fischer for his letter saying that he, and his collaborators,
were convinced that the unusual behavior of their differential cross
sections are a real effect. We thank Pr. B. Krusche for the numerical values
of $\eta$ meson photoproduction sent to us and Pr. H.P. Morsch for
stimulating discussions.

\begin{table}
\caption{pp$\to$pp$\eta$ reaction. Properties (in MeV) of the peaks extracted
using different backgrounds at $T_{p}$=1520~MeV, $\theta$=2$^{0}$. Insert (a)
corresponds to the result of the subtraction of the simulated background, 
insert (b) corresponds to the subtraction of the experimental background in both
parts of the $\eta$ meson window in the missing range (see text). R
indicates the ratio of the peak surface relative to surface of the 1564~MeV peak.}
\label{Table 1}
\end{table}
\begin{table}
\begin{tabular} [h]{c c c c c c c c c c c c}
 & & First peak & &I & &Second peak & &I & &Third peak & \\
Insert &mass&$\sigma(\Gamma)$&R &I &mass&$\sigma(\Gamma)$&R &I
&mass&$\sigma(\Gamma)$&R\\
\hline
(a)&1502&8&1.36&I&1519&8&0.57&I&1564&7.5&1\\
(b)&1502&8&1.43&I&1519&8&0.71&I&1563&7.5&1\\ 
\end{tabular}
\end{table}

\begin{table}
\caption{The pp$\to$pp$\eta$ reaction at $T_{p}$=1520~MeV, $\theta$=2$^{0}$.
Properties (in MeV) of the peaks extracted in different kinematical
conditions. The software cuts select
the $\eta$ meson missing mass region. Inserts (a), (b), and (c)
correspond to the result of the subtraction of the simulated background
for different kinematical conditions (see text).}
\label{Table 2}
\end{table}
\begin{table}
\begin{tabular} [h]{c c c c}
Insert & First peak & Second Peak & Third peak\\
\hline
(a)&M=1502\hspace*{5mm}$\sigma(\Gamma)$=8&M=1519\hspace*{5mm}$\sigma(\Gamma)$
=8&M=1564\hspace*{5mm}$\sigma(\Gamma)$=7.5\\
(b) & & &consistent with M=1564\hspace*{5mm}$\sigma(\Gamma)$=8\\
(c) & & & M=1564\hspace*{5mm}$\sigma(\Gamma)$=7.5\\
\hline
\end{tabular}
\end{table}

\begin{table}
\caption{Properties of the peaks extracted at
different angles from the pp$\to$pp$\pi^{0}$ reaction at
$T_{p}$=1520~MeV. Inserts (a) and (b) show the M$_{Xps}$ and
M$_{Xpf}$ data respectively. R denotes the ratio of peak surfaces (higher 
mass peak divided by the lower mass peak), where the two
peaks are extracted under the same kinematical conditions. S.D. is the
number of standard deviations of each extracted peak. The masses M and widths
$\sigma(\Gamma)$ are in MeV. All figures (except figure 14) show upper branch
events.}
\label{Table 3}
\end{table}
\begin{table}
\begin{tabular} [h]{c c c c c c c c c c c}
$\theta$&fig.&insert&branch&variable&M\hspace*{2.mm}($\sigma(\Gamma)$)&
&M\hspace*{2.mm}($\sigma(\Gamma)$) & &\\
deg.& & & & &(MeV)&S.D.&(MeV)&S.D.&R\\
\hline
0&16&a&M$_{\pi^{0}ps}$&-&-&-&-&-\\
 & & b&M$_{\pi^{0}pf}$&-&-&-&-&-\\
\hline
2&17&a&up&M$_{\pi^{0}ps}$&1479 (3.9)&5.7&1490.2 (2.0)&3.4&0.43\\
& &b&up&M$_{\pi^{0}pf}$&-&-&-&-&-\\
\hline
5&18&a&M$_{\pi^{0}ps}$&-&-&-&-&-\\
 & & b&M$_{\pi^{0}pf}$&-&-&-&-&-\\
\hline
9&19&a&up&M$_{\pi^{0}ps}$&1512.0 (2.8)&4.0&-&-&-\\
& &b&up&M$_{\pi^{0}pf}$&-&-&-&-&-\\
\hline
13&20&a&up&M$_{\pi^{0}ps}$&1513.7 (7.8)&9.1&-&-&-\\
&&b&up&M$_{\pi^{0}pf}$&1520.5 (3.5)&6.0&1531.9 (2.7)&8.2&1.54\\
\hline
17&21&a&up&M$_{\pi^{0}ps}$&1514.2 (9)&24&-&-&-\\
&&b&up&M$_{\pi^{0}pf}$&1520 (5)&12.6&1531.5 (4.7)&21.3&1.94\\
\hline
13&22&a&low&M$_{\pi^{0}pf}$&-&-&-&-&-\\
17&  &b&low&M$_{\pi^{0}pf}$&1520 (5.5)&23.4&1531 (4.7)&15.9&0.55\\
\end{tabular}
\end{table}

\begin{table}
\caption{Missing mass from the pp$\to$p$_{f}$p$_{s}\pi^{0}$ reaction
at $T_{p}$=1805~MeV, $\theta$=13$^{0}$ and 17$^{0}$.}
\label{Table 4}
\end{table}
\begin{table}
\begin{tabular} [h]{c c c c c c c}
fig.&insert&$\theta$&variable&branch&M\hspace*{2.mm}($\sigma(\Gamma)$)&M
\hspace*{2.mm}($\sigma(\Gamma)$)\\
\hline
23&(a) and (b)&13&M$_{\pi^{0}ps}$&up&1558 (5)&1582 (7.7)\\
 & & & & &1600 (7.7)&\\
24&(a) and (b)&13&M$_{\pi^{0}pf}$&up&&\\
\hline
6&(a) and (b)&17&M$_{\pi^{0}ps}$&up&1534.0 (8.0)&1597.5 (12.5)\\
25&(a) and (b)&17&M$_{\pi^{0}pf}$&up&1601.1 (4.0)&1622.6 (7.5)\\
26 &  &17&M$_{\pi^{0}pf}$ and M$_{\pi^{0}ps}$&up and down&1602
(6.4)&1620.5 (7.9)\\
\end{tabular}
\end{table}

\begin{table}
\caption{Missing mass from the pp$\to$p$_{f}$p$_{s}\eta$ reaction
at $T_{p}$=1520~MeV.
Properties of the peaks extracted at different angles.
 The software cuts select the $\eta$ meson missing mass
region. The masses and widths ($\sigma(\Gamma)$) are in MeV.
R denotes the peak surface relative to the peak surface at M=1564~MeV.
The masses and widths are in MeV.}
\label{Table 5}
\end{table}
\begin{table}
\begin{tabular} [h]{c c c c c c c c c c c c c c c}
Angle&figure&branch&variable&mass&$\sigma(\Gamma)$&R&I&mass&$\sigma(\Gamma)$&R&I&
mass&$\sigma(\Gamma)$&R\\
\hline
0$^{0}$&29&up&M$_{ps\eta}$&1500&8&1.37&I&1519&8&0.42&I&1564&7.5&1\\
2$^{0}$&7&up&M$_{ps\eta}$&1502&8&1.36&I&1519&8&0.57&I&1564&7.5&1\\
5$^{0}$&30&up&M$_{ps\eta}$&1502&8.5&0.09&I&1520.5&9&1.08&I&1567&8&1\\
9$^{0}$&31&up&M$_{ps\eta}$& & & &I& & & &I&1564.2&6&1\\
\hline
0$^{0}$&33&lo&M$_{pf\eta}$&1561.5&6.5&1&I&1580&6.5&0.68&&\\
2$^{0}$&34&lo&M$_{pf\eta}$&1562.4&6&1&1&1580.1&6&1.15&&\\
5$^{0}$&35&lo&M$_{pf\eta}$&1560.1&6.5&1&I&1576.2&6.1&1.08&&\\
9$^{0}$&36&lo&M$_{pf\eta}$&1553.1&6.3&0.35&I&1567.0&5.3&1.&&\\
\end{tabular}
\end{table}

\begin{table}
\caption{Missing mass from the pp$\to$p$_{f}$p$_{s}\eta$ reaction
at $T_{p}$=1805~MeV.
Properties of the peaks extracted at different angles.
 The software cuts select the $\eta$ meson missing mass
region. The masses and widths ($\sigma(\Gamma)$) are in MeV. R gives the
ratio of the first peak relative to the second peak.}
\label{Table 6}
\end{table}
\begin{table}
\begin{tabular} [h]{c c c c c c c} 
Insert&Angle &mass&$\sigma(\Gamma)$&R&mass&$\sigma(\Gamma)$\\
\hline
(a)&0.75$^{0}$&1642.5&6&0.71&1656.9&6\\
(b)&3.7$^{0}$&1642.3&7.5&0.89&1658.7&7.3\\
(c)&6.7$^{0}$&1644.2&7.8&1.45&1659.8&7.5\\
\hline
\end{tabular}
\end{table}

\begin{table}
\caption{Quantitative information concerning the narrow baryonic structures
extracted from the following reactions: pp$\to$ppX [1], pp$\to$p$\pi^{+}$n
[2], pp$\to$pp$\pi^{0}$ [3], and pp$\to$pp$\eta$ [4].
These informations are used in Fig.~45. The masses are in MeV, the angles
are in degrees.}
\label{Table 7}
\end{table}
\begin{table}
\begin{tabular} [h]{c c c c c c c c c c c c}
channel&Fig.&reaction&T&$\theta$&variable&branch&1st mass& 2nd mass& 3rd mass
& 4th mass\\
\hline
(a)&10&[1]&1520&0$^{0}$&M$_{psX}$&up&1535&&&\\
(b)&10&[1]&1520&2$^{0}$&M$_{psX}$&up&1482&1500&1535&1565\\
(c)&10&[1]&1520&5$^{0}$&M$_{psX}$&up&1476&1538&&\\
(d)&10&[1]&1520&9$^{0}$&M$_{psX}$&up&1535&1575&&\\
(e)&14&[1]&1805&0.75$^{0}$&M$_{psX}$&up&1635&1668&&\\
(f)&14&[1]&1805&3.7$^{0}$&M$_{psX}$&up&1637&1669&&\\
(g)&14&[1]&1805&6.7$^{0}$&M$_{psX}$&up&1667&&&\\
(h)&14&[1]&1805&0.75$^{0}$&M$_{pfX}$&up&1636&1666&&\\
(i)&14&[1]&1805&3.7$^{0}$&M$_{pfX}$&up&1637&&&\\
(j)&14&[1]&1805&6.7$^{0}$&M$_{pfX}$&up&1640&1659&&\\
(k)&14&[1]&1805&9$^{0}$&M$_{pfX}$&up&1641&&&\\
\hline
(l)&15&[2]&1520&0$^{0}$&M$_{n\pi^{+}}$&up&1505&1540&&\\
(m)&15&[2]&1520&2$^{0}$&M$_{n\pi^{+}}$&up&1508&1545&&\\
(n)&15&[2]&1520&5$^{0}$&M$_{n\pi^{+}}$&up&1506&(1540?)&&\\
(o)&15&[2]&1520&9$^{0}$&M$_{n\pi^{+}}$&up&1508&&&\\
(p)&15&[2]&1520&13$^{0}$&M$_{n\pi^{+}}$&up&1588&&&\\
(q)&15&[2]&1520&17$^{0}$&M$_{n\pi^{+}}$&up&(1550?)&&&\\
\hline
(r)&17&[3]&1520&2$^{0}$&M$_{ps\pi^{0}}$&up&1479&1490&&\\
(s)&19&[3]&1520&9$^{0}$&M$_{ps\pi^{0}}$&up&1512&&&\\
(t)&20&[3]&1520&13$^{0}$&M$_{ps\pi^{0}}$&up&1513.7&&&\\
(u)&20&[3]&1520&13$^{0}$&M$_{pf\pi^{0}}$&up&1520.5&1531.9&&\\
(v)&21&[3]&1520&17$^{0}$&M$_{ps\pi^{0}}$&up&1514.2&&&\\
(w)&21&[3]&1520&17$^{0}$&M$_{pf\pi^{0}}$&up&1520&1531.5&&\\
(x)&22&[3]&1520&17$^{0}$&M$_{pf\pi^{0}}$&lo&1520&1531&&\\
(y)&23&[3]&1805&13$^{0}$&M$_{ps\pi^{0}}$&up&1558&1582&1600.3&\\
(z)&24&[3]&1805&13$^{0}$&M$_{pf\pi^{0}}$&up&&&&\\
($\alpha$)&6&[3]&1805&17$^{0}$&M$_{ps\pi^{0}}$&up&1534&1597.5&&\\
($\beta$)&25&[3]&1805&17$^{0}$&M$_{pf\pi^{0}}$&up&1601.1&1622.6&&\\
($\eta$)&26&[3]&1805&17$^{0}$&M$_{p\pi^{0}}$&all&1599.5&1620.5&&\\
\hline
($\delta$)&29&[4]&1520&0$^{0}$&M$_{ps\eta}$&up&1500&1519&1564&\\
($\epsilon$)&7&[4]&1520&2$^{0}$&M$_{ps\eta}$&up&1502&1519.5&1564&\\
($\varphi$)&30&[4]&1520&5$^{0}$&M$_{ps\eta}$&up&1520.5&1567&&\\
($\gamma$)&31&[4]&1520&9$^{0}$&M$_{ps\eta}$&up&1564.2&&&\\
($\chi$)&33&[4]&1520&0$^{0}$&M$_{pf\eta}$&lo&1561.5&1580&&\\
($\iota$)&34&[4]&1520&2$^{0}$&M$_{pf\eta}$&lo&1562.4&1580.1&&\\
($\pi$)&35&[4]&1520&5$^{0}$&M$_{pf\eta}$&lo&1560.1&1576.2&&\\
($\kappa$)&36&[4]&1520&9$^{0}$&M$_{pf\eta}$&lo&1553.1&1567&&\\
($\lambda$)&43&[4]&1805&0.75$^{0}$&M$_{p\eta}$&all&1642.5&1656.9&&\\
($\mu$)&43&[4]&1805&3.7$^{0}$&M$_{p\eta}$&all&1642.3&1658.7&&\\
($\nu$)&43&[4]&1805&6.7$^{0}$&M$_{p\eta}$&all&1644.2&1659.8&&\\ 
\hline
\end{tabular}
\end{table}

\begin{table}
\caption{Masses (in MeV) of narrow exotic baryons observed in SPES3 data
and extracted from previous p($\alpha,\alpha'$)X spectra measured at SPES4
[24].}
\label{Table 8}
\end{table}
\begin{table}
\begin{tabular} [h]{c c c c c c c c c }
\hline
SPES3 mass&1479&1505&1517&1533&1542&(1554)&1564&1577\\
peak marker&(l)&(m)&(n)&(o)&(p)&(q)&(r)&(s)\\
SPES4 mass 2$^{0}$&1477&1507&1517&1530&1544&1557&1569&1580\\
\hline
\end{tabular}
\end{table}

\begin{table}
\caption{Total cross-section of $\eta$ meson photoproduction on the proton.
Measurements performed at MAMI [31] [32]. Small peaks (a1), (a2), and (a3) 
are extracted (see Fig.~49).}
\label{Table 9}
\end{table}
\begin{table}
\begin{tabular} [h]{c c c c}
\hline
peak&$E_{\gamma}$~(MeV)&$\surd(s)$~(MeV)&SPES3 mass (MeV)\\
(a1)&739.8&1506&1505\\
(a2)&757.2&1517&1517\\
(a3)&786.0&1534.7&1533\\
\hline
\end{tabular}
\end{table}

\begin{figure}
\caption{Spes3 spectrometer and the associated detection.}
\label{fig1}
\end{figure}

\begin{figure}
\caption{pp$\to$ppX reaction at T$_{p}$=1520~MeV and $\theta$=2$^{0}$.
Scatterplot of fast proton momenta p$_{f}$ versus the invariant mass
M$_{Xps}$ and slow proton momenta p$_{s}$ versus M$_{Xpf}$.}
\label{fig2}
\end{figure}

\begin{figure}
\caption{Number of events of the missing mass of the pp$\to p_{f}p_{s}$X
scattering. Inserts (a) and (b) show T$_{p}$=1520~MeV data for the upper
branch $p_{f}\ge~p_{f}$limit, respectively at $0^{0}$ and $17^{0}$.
Inserts (c) and (d) show T$_{p}$=1805~MeV data for the upper
branch $p_{s}\ge~p_{s}$limit, respectively at $0.75^{0}$ and $17^{0}$.}
\label{fig3}
\end{figure}
 
\begin{figure}
\caption{pp$\to$pp$\pi^{0}$ reaction at T$_{p}$=1520~MeV, $\theta$=5$^{0}$.
Inserts (a) ((b)) shows the number of events versus the momenta of
the fast (slow) proton, for the upper branch of 
M$_{\pi^{0}ps}$(M$_{\pi^{0}pf}$). The lower branches are empty due to the
M$_{X}\le$250~MeV software cut. Insert (c) shows the missing mass under
the same conditions as in insert (a). Full points are data and empty points
are simulated.}
\label{fig4}
\end{figure}

\begin{figure}
\caption{Same caption as caption of fig.~4, but for $\theta$=9$^{0}$.}
\label{fig5}
\end{figure}

\begin{figure}
\caption{(a) The number of events of the upper branch of the
M$_{\pi^{0}ps}$ invariant mass from the pp$\to$pp$\pi^{0}$ reaction at
T$_{p}$=1805~MeV and
$\theta$=17$^{0}$. Full points are the measured data; empty points are
the corresponding normalized simulated events (see text). (b) 
The difference between data and simulation, representing correlated
invariant masses from narrow baryonic resonances.}
\label{fig6}
\end{figure}

\begin{figure}
\caption{pp$\to p_{f}p_{s}\eta$ scattering at T$_{p}$=1520~MeV and
$\theta$=2$^{0}$. Number of events versus $M_{ps\eta}$ selected by a
cut on $p_{f}$ in order to retain only the upper branch events.  In insert (a),
the full points are the data and the empty points are the normalized
simulated background. In insert (b), the difference between data and
background is presented and is fitted by three gaussians (see table 1).}
\label{fig7}
\end{figure}

\begin{figure}
\caption{Data minus normalized simulations for the uncorrelated events
at T$_{p}$=1520~MeV, $\theta=2^{0}$.
The three inserts (a), (b), and (c) exhibit respectively the number of
events versus M$_{Xps}$ when p$_{f}\ge~p_{f}$limit (upper branch), versus
M$_{Xpf}$ when p$_{s}\ge~p_{s}$limit (upper branch), and versus M$_{Xpf}$ when
p$_{s}\le~p_{s}$limit (lower branch).}
\label{fig8}
\end{figure}

\begin{figure}
\caption{Same caption as the one of Fig.~7, except that the normalized
background is the physical background obtained by selecting lower and upper
part of the missing mass spectrum around the $\eta$ meson mass.}
\label{fig9}
\end{figure}

\begin{figure}
\caption{Upper branch of the pp$\to$ppX cross sections at
$T_{p}$=1520~MeV versus M$_{psX}$.
In order to appear clearly in the same figure
the data in insert (a) are normalized by factors: 1.7, 2.2, 3.5, and
2.5 for angles $0^{0}$, $2^{0}$, $5^{0}$, and $9^{0}$ respectively.
The right-hand part of the figure shows the $\theta=2^{0}$ data with
different binnings. The four inserts
(a), (b), (c), and (d) correspond respectively to the following binnings:
 0.8, 3.2, 19.2, and 32~MeV/channel.}
\label{fig10}
\end{figure}

\begin{figure}
\caption{Lower branch of the pp$\to$ppX differential cross section at
$T_{p}$=1520~MeV versus M$_{psX}$.
In order to appear clearly in the same figure
the data are normalized by the following factors: 1.5, 1.5, 0.8, 0.3, 0.1,
and 0.1 for
angles $0^{0}$, $2^{0}$, $5^{0}$, $9^{0}$, $13^{0}$, and $17^{0}$,
respectively.}
\label{fig11}
\end{figure}

\begin{figure}
\caption{Upper branch of the pp$\to$ppX differential cross section
at $T_{p}$=1520~MeV versus M$_{pfX}$.
In order to appear clearly in the same figure
the data are normalized by the following factors: 1.0, 0.4, 0.5, 0.3, 0.1,
and 0.1 for
angles $0^{0}$, $2^{0}$, $5^{0}$, $9^{0}$, $13^{0}$, and $17^{0}$,
respectively.}
\label{fig12}
\end{figure}
 
\begin{figure}
\caption{Lower branch of the pp$\to$ppX cross sections at
$T_{p}$=1520~MeV versus M$_{pfX}$.
In order to appear clearly in the same figure
the data are normalized by the following factors: 1, 0.4, 0.6, 0.5, 0.4,
and 0.7 for
angles $0^{0}$, $2^{0}$, $5^{0}$, $9^{0}$, $13^{0}$, and $17^{0}$,
respectively.}
\label{fig13}
\end{figure}

\begin{figure}
\caption{Cross sections of the upper branch of the pp$\to$ppX reaction
at $T_{p}$=1805~MeV. The left-hand part of the figure shows the results from
the upper branch versus
M$_{psX}$. In order to appear clearly in the same figure
the data are normalized by the following factors: 3.6, 7.5, 10.0, 7.5 and 10.0
for angles $0.75^{0}$, $3.7^{0}$, $6.7^{0}$, $9^{0}$, and $13^{0}$,
respectively. The right-hand part of the figure shows the results from the upper
branch versus M$_{pfX}$. In order to appear clearly in the same figure
the data are normalized by the following factors: 1, 2.5, 4, 4, and 5 for
 angles $0.75^{0}$, $3.7^{0}$, $6.7^{0}$, $9^{0}$, and $13^{0}$,
respectively.}
\label{fig14}
\end{figure}

\begin{figure}
\caption{The left-hand part of the figure shows the differential cross
sections of the $pp\to~p\pi^{+}n$ reaction versus $M_{\pi^{+}n}$, at
forward angles and at $T_{p}$=1520~MeV.
In order to appear clearly in the same figure, the data, going from top to
bottom, are normalized by factors: 1.0 at
$0^{0}$ (empty squares), 1.5 at $2^{0}$ (full circles), 4.4 at $5^{0}$
(full triangles), and 12. at $9^{0}$ (empty stars).
The right-hand part of the figure shows the same differential cross sections
at $T_{p}$=1805~MeV, at the two largest measured angles. In order to appear
clearly in the same figure, the data are normalized by
factors: 1.0 at $13^{0}$ (full triangles) and 2.0 at $17^{0}$ (empty
triangles).}
\label{fig15}
\end{figure}

\begin{figure}
\caption{pp$\to$pp$\pi^{0}$ reaction at $T_{p}$=1520~MeV and
$\theta$=0$^{0}$. Both inserts show the number of events of the upper branch
versus M$_{\pi^{0}ps}$ (insert (a)), and versus M$_{\pi^{0}pf}$ (insert (b)).
Full circles show the data and empty circles show
the normalized simulated background (see text for more details).}
\label{fig16}
\end{figure}

\begin{figure}
\caption{pp$\to$pp$\pi^{0}$ reaction at $T_{p}$=1520~MeV and $\theta$=2$^{0}$.
Both inserts show the number of events of the upper branch
versus M$_{\pi^{0}ps}$ (insert (a)), and versus M$_{\pi^{0}pf}$ (insert (b)).
Full circles show the data and empty circles show
the normalized simulated background.
Full triangles show the difference between data and simulation. The curves show
two gaussians tentatively extracted from the full triangles spectrum.
(see text for more details).}
\label{fig17}
\end{figure}

\begin{figure}
\caption{Same caption as for Fig.~17, but at $\theta$=5$^{0}$.}
\label{fig18}
\end{figure}

\begin{figure}
\caption{Same caption as for Fig.~10, but at $\theta$=9$^{0}$.}
\label{fig19}
\end{figure}

\begin{figure}
\caption{Same caption as for Fig.~10, but at $\theta$=13$^{0}$.}
\label{fig20}
\end{figure}

\begin{figure}
\caption{Same caption as for Fig.~10, but at $\theta$=17$^{0}$.}
\label{fig21}
\end{figure}

\begin{figure}
\caption{pp$\to$pp$\pi^{0}$ reaction at T$_{p}$=1520~MeV and
$\theta$=13$^{0}$ (insert (a)) and $\theta$=17$^{0}$ (insert (b)).
Number of events of the lower branch versus M$_{pf\pi^{0}}$. Full
circles show the data, empty circles show
the normalized simulated background, and
full triangles show the difference between measured and simulated data.
See text for more details, and table~3 for quantitative information.}
\label{fig22}
\end{figure}

\begin{figure}
\caption{pp$\to$pp$\pi^{0}$ reaction at $T_{p}$=1805~MeV and
$\theta$=13$^{0}$. Number of events of the upper branch, versus
M$_{ps\pi^{0}}$. Insert
(a) shows the data (full points) and simulated uncorrelated events
(empty points). Insert (b) shows the difference between both
(full triangles) and an attempt to extract three gaussians from the
spectrum.}
\label{fig23}
\end{figure}

\begin{figure}
\caption{Same caption as for Fig.~15, but versus M$_{pf\pi^{0}}$.
Two different possible descriptions of the data, with one or two gaussians,
are shown, although the description with two gaussians is preferred
(see text).} 
\label{fig24}
\end{figure}

\begin{figure}
\caption{Same caption as for Fig.~3, but versus M$_{pf\pi^{0}}$.}           
\label{fig25}
\end{figure}

\begin{figure}
\caption{The pp$\to$pp$\pi^{0}$ reaction at $T_{p}$=1805~MeV and
$\theta$=17$^{0}$. Addition of data from both upper branches. The
quantitative information is given in table~4.}
\label{fig26}
\end{figure}

\begin{figure}
\caption{The upper branch data corresponding to the pp$\to$pp$\pi^{0}$
reaction at $T_{p}$=1805~MeV and
$\theta$=17$^{0}$. Insert (a) shows the scatterplot of the fast versus the
slow detected proton momenta. Insert (b) shows the same scatterplot after
a software selection vetoing the range 1600$\le$M$_{p\pi^{0}}$$\le$1610~MeV.
Insert (c) shows the scatterplot of these data corresponding to the range
1600$\le$M$_{p\pi^{0}}$$\le$1610~MeV.}
\label{fig27}
\end{figure}

\begin{figure}
\caption{The pp$\to$ppX reaction at $T_{p}$=1520~MeV and
$\theta$=0$^{0}$. Inserts (a), (b), (c), and (d) show the missing mass
M$_{X}$ for p$_{s}\ge$p$_{s}$limit, p$_{f}\ge$p$_{f}$limit,
p$_{s}\le$p$_{s}$limit, and p$_{f}\le$p$_{f}$limit respectively.}
\label{fig28}
\end{figure}

\begin{figure}
\caption{The pp$\to$pp$\eta$ data, selected by software cuts, at
T$_{p}$=1520~MeV and $\theta$=0$^{0}$. Insert (a) shows the data of the
upper branch (full points), and the corresponding normalized simulated
background (empty points), versus M$_{ps\eta}$.
Insert (b) shows the data minus background spectrum, and the extracted
gaussians.}
\label{fig29}
\end{figure}

\begin{figure}
\caption{Same caption as in Fig.~20, but for $\theta$=5$^{0}$.}
\label{fig30}
\end{figure}

\begin{figure}
\caption{Same caption as in Fig.~20, but for $\theta$=9$^{0}$.}
\label{fig31}
\end{figure}

\begin{figure}
\caption{pp$\to$pp$\eta$ data selected by software cuts at T$_{p}$=1520~MeV.
Relative intensities R, versus the spectrometer angle (see text). In
spectra versus
M$_{ps\eta}$, full circles show the ratio of the M$\approx$1505~MeV
peak over the M$\approx$1564~MeV peak, full triangles show the ratio of the
M$\approx$1517~MeV peak over the M=1564~MeV peak. In the spectra versus
M$_{pf\eta}$,
full stars show the ratio of the M$\approx$1577~MeV peak over the
M$\approx$1564~MeV peak.}
\label{fig32}
\end{figure}

\begin{figure}
\caption{Number of events of the lower branch of the pp$\to$pp$\eta$
reaction, selected by software cuts, at
T$_{p}$=1520~MeV and $\theta$=0$^{0}$. Insert (a) shows the data (full points)
and the normalized simulated background (empty points) versus M$_{pf\eta}$.
Insert (b) shows the data minus background, and the extracted gaussians.}
\label{fig33}
\end{figure}

\begin{figure}
\caption{Same caption as in Fig.~23, but for $\theta$=2$^{0}$.}
\label{fig34}
\end{figure}

\begin{figure}
\caption{Same caption as in Fig.~23, but for $\theta$=5$^{0}$.}
\label{fig35}
\end{figure}

\begin{figure}
\caption{Same caption as in Fig.~23, but for $\theta$=9$^{0}$.}
\label{fig36}
\end{figure}

\begin{figure}
\caption{The pp$\to$pp$\eta$ reaction at $T_{p}$=1805~MeV and
$\theta$=0.75$^{0}$.
Insert (a) and (c) show the missing mass events corresponding to the upper
branch of M$_{Xpf}$ and to the upper branch of the M$_{Xps}$ spectra. Inserts
(b) and (d) show the corresponding invariant masses. Full points are data
without cuts on the missing mass, and empty points show the number of events
after applying software cuts to retain the $\eta$ meson in the missing mass.}
\label{fig37}
\end{figure}

\begin{figure}
\caption{The pp$\to$pp$\eta$ reaction at $T_{p}$=1805~MeV and
$\theta$=0.75$^{0}$. Inserts (a), (b), and (c) show both the two upper branch
data, both lower branch data, and data from all four branches respectively.
The simulated background is reported by two straight lines.}
\label{fig38}
\end{figure}

\begin{figure}
\caption{The pp$\to$pp$\eta$ reaction at $T_{p}$=1805~MeV. Insert (a) shows the
simulated events at $\theta$=0.75$^{0}$, from the upper branch of
M$_{pf\eta}$ (empty symbols) and from the upper branch of M$_{ps\eta}$
(filled symbols). This simulated data is fitted by two
straight lines. Insert (b) shows the same results for $\theta$=3.7$^{0}$.}
\label{fig39}
\end{figure}

\begin{figure}
\caption{Same caption as in Fig.~27 but for $\theta$=3.7$^{0}$.}
\label{fig40}
\end{figure}

\begin{figure}
\caption{Same caption as Fig.~28, but for $\theta$=6.7$^{0}$.}
\label{fig41}
\end{figure}

\begin{figure}
\caption{Same caption as in Fig.~29, but for $\theta$=6.7$^{0}$.}
\label{fig42}
\end{figure}

\begin{figure}
\caption{The pp$\to$pp$\eta$ reaction at $T_{p}$=1805~MeV. The points show the
total data minus the normalized uncorrelated background simulated data.
Inserts (a), (b), and (c) are for $\theta$=0.75$^{0}$, 3.7$^{0}$, and
6.7$^{0}$ respectively.}
\label{fig43}
\end{figure}

\begin{figure}
\caption{The pp$\to$pp$\eta$ reaction at $T_{p}$=1805~MeV. The ratio R 
shows the relative excitation between the
M$\approx$1639~MeV and the M$\approx$1659~MeV peaks at three forward angles,
where both peaks were extracted.}
\label{fig44}
\end{figure}

\begin{figure}
\caption{Masses (in MeV) of the narrow structures extracted from pp$\to$ppX
and pp$\to$p$\pi^{+}$n reactions in the mass range 1470$\le$M$\le$1680~MeV.
Full (empty) circles correspond to well
(less well) defined peaks. Each column correspond to a particular
experimental situation (reaction, incident energy, angle, upper or lower
branch, and invariant mass reconstructed using the slow (fast) proton).
The mass ranges studied differ with different reactions; they are indicated
by a dashed area. The mass values of these areas are defined in the following
way: the mean value
of several masses of narrow peaks is kept as being a narrow peak mass, and
the area is drawn with $\pm$3~MeV range. The quantitative information is
given in table~VII.}
\label{fig45}
\end{figure}

\begin{figure}
\caption{The figure shows the disintegration channels of the narrow baryonic
resonances experimentally observed, and an attempt to associate them with
broad PDG resonances.}
\label{fig46}
\end{figure}

\begin{figure}
\caption{The p($\alpha,\alpha'$)X spectrum at T$_{\alpha}$=4.2~GeV,
$\theta$=2$^{0}$, measured at SPES4 (Saturne) [24]. Only the
part of the spectrum corresponding to the baryonic mass range studied here:
1470$\le$M$\le$1680~MeV is shown.}
\label{fig47}
\end{figure}

\begin{figure}
\caption{Comparison of narrow mass structures observed in the SPES3 data
with those extracted from the p($\alpha,\alpha'$)X spectrum at
T$_{\alpha}$=4.2~GeV, $\theta$=2$^{0}$, measured at SPES4 (Saturne)
[24]. Only part of the spectrum, corresponding to the baryonic
mass range studied here: 1470$\le$M$\le$1680~MeV, is shown.}
\label{fig48}
\end{figure}

\begin{figure}
\caption{Near threshold photoproduction of the $\eta$ meson on the proton.
The data were taken at MAMI (Mainz) [31] [32]. The quantitative information
of the extracted peaks is given in table~IX.}
\label{fig49}
\end{figure}

\begin{figure}
\caption{Differential cross section at $\theta=65^{0}$ of the
$\gamma$p$\to\pi^{+}$n reaction measured at Bonn [36].}
\label{fig50}
\end{figure}

\end{document}